\RequirePackage{amsmath}

\documentclass{aip-cp}

\usepackage[numbers]{natbib}
\usepackage{rotating}
\usepackage{graphicx}

\begin{document}


\title{Recent Results from the Daya Bay Neutrino Experiment}

\author{Wei Tang \\ {(On behalf of the Daya Bay Collaboration)}}

\affil{Physics Department, Brookhaven National Laboratory, Upton, New York 11973, USA}

\maketitle

\begin{abstract}
The Daya Bay neutrino experiment has recently updated the oscillation analysis results with 621 days of data in 2015, which has 3.6 times more statistics than the previous publication in 2014. The relative $\bar{\nu}_{e}$ rate and spectrum measurement between the near and far detectors yielded the best fit values of $\sin^{2}2\theta_{13}$ = 0.084 $\pm$ 0.005 and $|\Delta{}m_{ee}^{2}|$ = (2.42 $\pm$ 0.11) $\times$ 10$^{-3}$ eV$^{2}$. This is currently the most precise measurement of $\sin^{2}2\theta_{13}$ in the world. The measurement of $|\Delta{}m_{ee}^{2}|$ also has a precision that is comparable to the measurements from MINOS and T2K experiments in 2014. Daya Bay also performed several other analyses such as the search for the light sterile neutrino in the 3+1 neutrino framework, and the measurements of the absolute reactor anti-neutrino flux and spectrum.  
\end{abstract}

\section{INTRODUCTION}

Neutrino was proposed by W. Pauli in 1930 to explain the continuous $\beta$ spectrum that had been observed. However, its detection was not realized until 26 years later: in 1956 neutrino was first detected in the Savannah River Experiment led by F. Reines and C. Cowan. Today we know that there are three types or flavors of active neutrinos ($\nu_{e}$, $\nu_{\mu}$ and $\nu_{\tau}$), and we also know neutrinos have masses (with mass eigenstates $\nu_{1}$, $\nu_{2}$ and $\nu_{3}$). Nevertheless, the neutrino flavor eigenstates are not their mass eigenstates. From the quantum mechanics, we know that neutrinos oscillate: the flavor components of the neutrinos will change as they propagate. 

The neutrino flavor and mass eigenstates can be bridged as
\begin{equation}
|\nu_{\alpha}> = \sum_{i=1}^{3}U_{\alpha,i}|\nu_{i}>,
\label{eq:nu_mix}
\end{equation}
where $\nu_{\alpha}$ and $\nu_{i}$ represent neutrino flavor and mass eigenstates, respectively. U is a 3 $\times$ 3 unitary matrix called Pontecorvo-Maki-Nakagawa-Sakata (PMNS) matrix that has the form of 
\begin{equation}  
U_{PMNS} = \begin{pmatrix}  1 & 0 & 0 \\ 0 & cos\theta_{23} & sin\theta_{23} \\ 0 & -sin\theta_{23}& cos\theta_{23}  \end{pmatrix}  \begin{pmatrix} \cos\theta_{13} & 0 & e^{-i\delta}\sin\theta_{13} \\ 0 & 1 & 0 \\ -e^{i\delta}\sin\theta_{13} & 0 & \cos\theta_{13}  \end{pmatrix}  \begin{pmatrix} \cos\theta_{12} & \sin\theta_{12} & 0 \\ -\sin\theta_{12} & \cos\theta_{12} & 0 \\ 0 & 0 & 1  \end{pmatrix}.
\label{eq:pmns}
\end{equation}
The PMNS matrix contains three neutrino mixing angles ($\theta_{12}$, $\theta_{13}$ and $\theta_{23}$) and one phase ($\delta$), and the $\delta$ could lead to the CP violation. For the Daya Bay Experiment, we focused on the measurement of the $\theta_{13}$ neutrino mixing angle using electron anti-neutrinos ($\bar{\nu}_{e}$) produced from reactors. 

The $\theta_{13}$ can be extracted by measuring the $\bar{\nu}_{e}$ survial probability after they propagate for certain distance, which has the form of
\begin{equation}
P_{ee} = P_{\bar{\nu}_{e} \to \bar{\nu}_{e} } = 1 - sin^{2}2\theta_{13}sin^{2}(\frac{\Delta{}m^{2}_{ee}L}{4E}) - cos^{4}\theta_{13}sin^{2}2\theta_{12}sin^{2}(\frac{\Delta{}m^{2}_{21}L}{4E}),
\label{eq:osc}
\end{equation}
\begin{equation}
sin^{2}(\frac{\Delta{}m^{2}_{ee}L}{4E}) \approx cos^{2}\theta_{12}sin^{2}(\frac{\Delta{}m^{2}_{31}L}{4E}) + sin^{2}\theta_{12}sin^{2}(\frac{\Delta{}m^{2}_{32}L}{4E}),
\label{eq:dee}
\end{equation}
where $E$ is the neutrino energy, $L$ is the neutrino propagation distance, and $\Delta{}m^{2}_{ij}$ = $m_{i}^{2}$ - $m_{j}^{2}$ $(i,j = 1,2,3)$.

Reactors mainly produce electron anti-neutrinos through the fissions of four isotopes ($^{235}U$, $^{238}U$, $^{239}Pu$ and $^{241}Pu$). On average, each fission can produce $\sim$ 6 electron anti-neutrinos and release $\sim$ 200 MeV thermal energy. Hence for a 1 GigaWatts (GW) thermal power reactor, it produces $\sim$2 $\times$ 10$^{20}$ electron anti-neutrinos per second with most neutrino energies below 8 MeV.

The detection of reactor anti-neutrinos is through the Inverse Beta Decay (IBD): $\bar{\nu}_{e}$ + p $\to$ e$^{+}$ + n that requires the $\bar{\nu}_{e}$'s energy larger than 1.8 MeV. The ionization and annihilation of e$^{+}$ in liquid scintillator forms a prompt signal, while the neutron will first get thermalized and then be captured by some nucleus and form a delayed signal. In the Daya Bay experiment, Gd is used to capture IBD neutrons. When a neutron is captured by Gd, it will release several gamma rays with total energy $\sim$ 8 MeV. The mean time difference between prompt and delayed signals is $\sim$ 30 $\mu$s. The coincidence of the prompt and the delayed signals forms a distinctive signature for IBD.    

In March, 2012, Daya Bay discovered a non-zero $\sin^{2}2\theta_{13}$ with 5.2 $\sigma$ significance using 55 days of data\citep{man:dyb_55days} and provided the most precise measurement of this parameter at that time. With 217 days of data, Daya Bay updated the results in 2014 with a relative $\bar{\nu}_{e}$ rate and spectrum analysis to measure both the $\sin^{2}2\theta_{13}$ and the oscillation frequency $|\Delta{}m^{2}_{ee}|$ \cite{man:dyb_217days}. In this article, we report the most recent results from Daya Bay using 621 days of data\cite{man:dyb_621days}, which has 3.6 more times of statistics than our previous published results\cite{man:dyb_217days}.

\section{THE DAYA BAY NEUTRINO EXPERIMENT}
The Daya Bay neutrino experiment is located inside the Daya Bay Nuclear Power Plant facility that is in the suburb of Shenzhen, Guangdong province, China. The Daya Bay experiment detects the electron anti-neutrinos produced from six 2.9 GW thermal power commericial reactors, Daya Bay nuclear power plants (two reactors), Ling Ao I nuclear power plants (two reactors) and Ling Ao II nuclear power plants (two reactors),  using eight functional identical Anti-neutrino Detectors (ADs). The ADs are placed in three underground experimental halls, two near Halls (EH1 and EH2) and one far Hall (EH3). EH1, which hosts two ADs, is 363 m away from Daya Bay nuclear power plants and has an overburden of 250 meter water equivalent (m.w.e.). EH2, with 265 m.w.e. overburden, also has two ADs and is about 500 m from Ling Ao I and II nuclear power plants. The other four ADs are located in EH3, which has an overburden of 860 m.w.e. and is 1615 and 1958 m from Ling Ao I and Daya Bay nuclear power plants, respectively. 

The AD used in the Daya Bay experiment is a 3-zone cylindrical shaped detector. The inner most part is a 3 $\times$ 3 m dimension Inner Acrylic Vessel (IAV) that holds 20 tons of Gd doped Liquid Scintilltor (Gd-LS) which is used as the detection target. The IAV is surrounded by 20 tons of none doped Liquid Scintillator (LS) that is used to capture the $\gamma$ rays leaked from Gd-LS region. The LS is holded by a 4 $\times$ 4 m Outer Acrylic Vessel (OAV). The outmost region is a 5 $\times$ 5 m Stainless Steel Vessel (SSV) which contains 40 tons of Mineral Oil (MO) to shield the Gd-LS and LS regions from the radioactivities produced by PMT glasses and the stainless steel. Each AD contains 192 8-inch PMTs that mounted on the side wall of SSV. Two optical reflectors are placed at the top and the bottom of SSV to increase the effective detector photo coverage to 12\%. 160 photoelectrons are collected per MeV on average which leads to the detector energy resolution of $\sim$8\%/$\sqrt{E(\mbox{MeV})}$.   

Three Automated Calibration Units (ACUs) are mounted on the top of SSV lid which are used to calibrate the detector Gd-LS center and edge as well as LS regions along three vertical axes. Each ACU contains a LED diffuser ball that is used to calibrate the PMT gain and timing, and three calibration sources: $^{68}$Ge, $^{60}$Co and $^{241}$Am$^{13}$C for $\gamma$ rays and neutron. The calibration is performed weekly to all 8 ADs simultaneously. Besides weekly calibration using ACU sources, during the summer of 2012 a special calibration campaign was performed on EH1's two ADs using $^{137}$Cs, $^{54}$Mn, $^{40}$K, $^{241}$Am$^{9}$Be and $^{239}$Pu$^{13}$C sources. In addition, a manual calibration system (MCS) was installed on EH1-AD1 during this period, which allows for the full volume calibration of the detector. Also, $\gamma$ and $\alpha$ peaks that from the radioactivities of the LS ($^{40}$K,$^{208}$Tl, $^{212}$Po, $^{214}$Po and $^{215}$Po) and neutron captured on H, C and Fe are included in the calibration. Overall the relative energy scale, the difference of the reconstructed energies between ADs, is less than 0.2\% for all sources.   

The ADs in each experimental hall are submerged in a purified water pool with at least 2.5 m of water coverage in all directions. The water pool is optically separated to 2-zone: Inner Water Pool (IWP) and Outer Water Pool (OWP), by a layer of Tyvek. The water pool is used as a Cherenkov detetor to veto cosmic muons as well as to shield the ADs from the radioactivities that come from the surrounding rock. Each water pool is covered with 4-layer Resistive Plate Chambers (RPCs) that can provide additional muon tagging information.  

The detector energy nonlinearity, or relative scintillator response to different energies of different types of particles (such as e$^{+}$, e$^{-}$ and $\gamma$), is important to the interpretation of the observed prompt energy spectra. There are two major sources that cause the detector nonlinearity response between the true energy and the reconstructed energies of a particle: the scintillator nonlinerity and the PMT read out electronics nonlinearity. The scintillator nonlinearity is due to the scintillator quenching effect, which can be described by Birk's law, and the Cherenkov light emission. The electronic nonlinearity is due to the interaction of the scintillation light time profile and the charge measurement scheme of the front-end electronics, which is modeled with Monte Carlo and measurement from single channel FADC in the Daya Bay experiment. The detector nonlinearity is correlated among all ADs, and is constrained to be $\sim$ 1\% in the Daya Bay experiment. For more information about Daya Bay experiment and its detector system, please check \cite{man:dyb_cpc,man:dyb_nim}. 

The Daya Bay experiment started to take data from December 2011 with 6-AD configuration, two ADs at EH1, one AD at EH2 and three ADs at EH3, and this period last for 217 days until July 2012. In the summer of 2012, two new ADs were installed: one at EH2 and the other at EH3. From then on, the Daya Bay experiment is running with full 8-AD configuration. In the following sections, we will present the recent results from Daya Bay using 621 days of data: 217 days data with 6-AD configuration and 404 days data with 8-AD configuration.

\section{IBD EVENTS SELECTION AND BACKGROUND}

To remove the irrelevant events and select the real IBD events that caused by the neutrinos, several cuts were applied to our IBD candidates. First of all, we removed the events that caused by PMT flashers: the light spontaneously emitted from PMT. Then we required the prompt and delayed energies of the IBD candidates to fall into the ranges of 0.7-12 MeV and 6-12 MeV, respectively. The prompt and delayed signals were also required to form a coincidence within a time window of 1-200 $\mu$s. To remove the events that caused by cosmic muons, muon veto cuts were applied. Events were removed if the delayed signal occurs within 600 $\mu$s after a water pool muon was detected, or the delayed signal happens within 1 ms after a reconstructed signal with energy $>$ 20 MeV in AD, or the delayed signal within 1 s after a reconstructed signal with energy $>$ 2.5 GeV in AD. Finally, to further remove ambiguity, a multiplicity cut is applied to select only a single candidate pairs within each cut window. The cut efficiency for the above mentioned cuts is 80.6\% $\pm$ 2.1\% $\pm$ 0.2\%, where the 2.1\% and 0.2\% are corresponding to the correlated and uncorrelated uncertainties. During the 621 days of data taking, we totally collected more than 1 million IBD events with $\sim$150 K events at far ADs. The time dependent IBD rates are strongly correlated with the reactor $\bar{\nu}_{e}$ flux expectation.     

The Daya Bay experiment is a low background experiment. There are five major backgrounds in the Daya Bay experiment that could mimic the IBD. The largest background is accidental background, which accounts for 2.3\% and 1.4\% of the candidates in the far and near halls, respectively. However, the accidental background can be statistically removed with high precision. The second and third backgrouns are from $^{9}$Li/$^{8}$He and fast neutrons that are generated from the cosmic muons account for 0.4\% and 0.1\% at both halls, respectively. The fourth background is due to the natural radioactivities within the liquid scintillator, which accounts for 0.01\% and 0.1\% at near and far halls. The last background is due to our $^{241}$Am$^{13}$C calibration source, which is largely reduced during the 8-AD configuration period due to the removal of off-center $^{241}$Am$^{13}$C sources at EH3 during the summer of 2012. The total background amount to 2\% and 3\% of the IBD candidates for the near and far halls.

\section{RECENT RESULTS FROM THE DAYA BAY EXPERIMENT}

In this section, we present the recent results from Daya Bay neutrino experiment that include the update of the measurement of $\sin^{2}2\theta_{13}$  and $|\Delta{}m_{ee}^{2}|$ values from the oscillation analysis using 621 days of data\cite{man:dyb_621days}, the light sterile neutrino searching results in the 3+1 neutrino framework\cite{man:dyb_sterile} and the absolute reactor $\bar{\nu}_{e}$ flux and positron spectrum measurement\cite{man:dyb_absolute} using 217 days of data.

\begin{figure}[h]
  \centerline{\includegraphics[width=380pt]{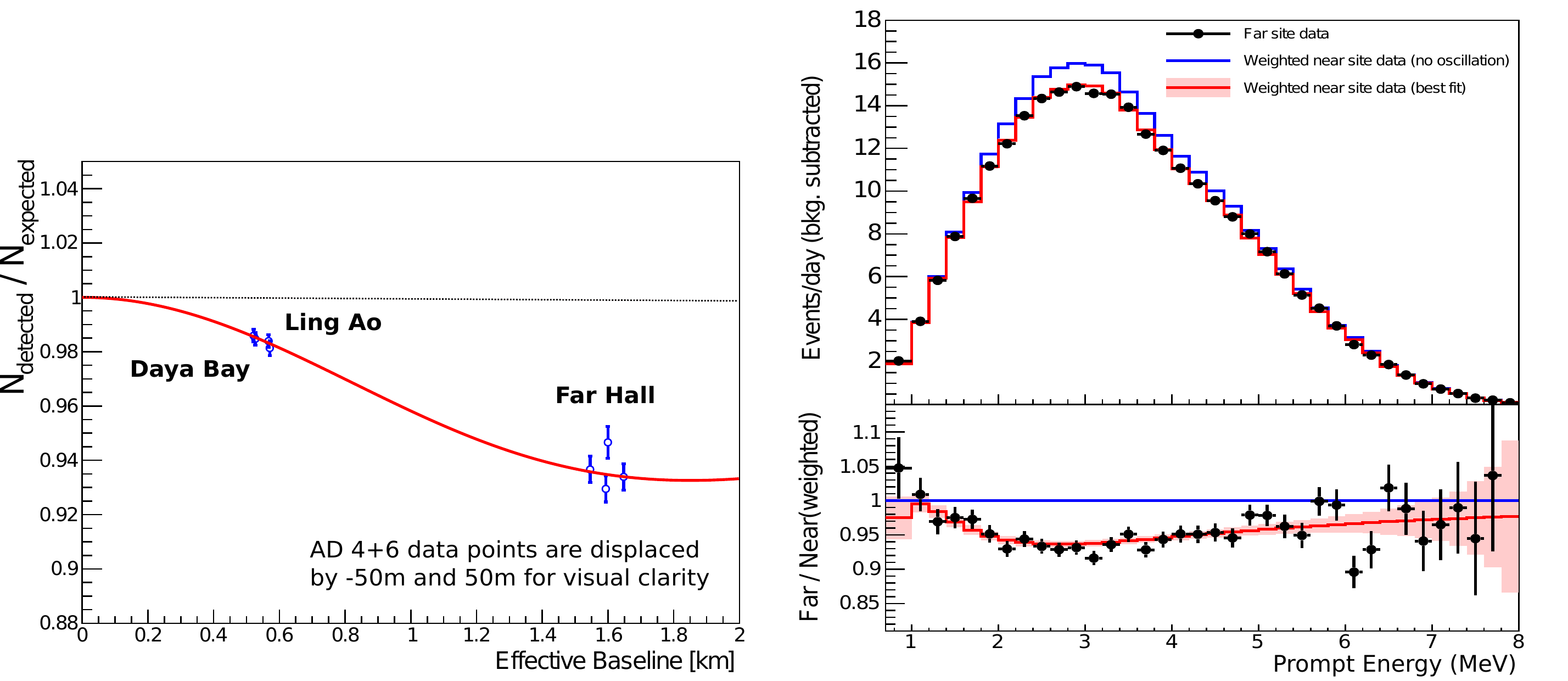}}
\caption{(Color online) Left: The ratio of the detected to expected $\bar{\nu}_{e}$ events that assuming no oscillation versus the effective baseline for each AD. The open dots represent the ratio for each AD, and the curve shows the best fit. Two far AD dots are displaced $\pm$ 50 m respectively on purpose to give a better view. Right: The top panel plot shows the measured prompt spectrum with background subtracted (solid dots) at far ADs compared to the expected prompt spectra based on the near ADs' data with (lower curve) and without (higher curve) oscillation interpretation. The bottom panel plot shows the ratio of the measured prompt spectrum at far ADs to the weighted near ADs' spectrum without oscillation, and the curve shows the best fit.}
\label{fig:osc_rate}
\end{figure}

The oscillation analysis is performed by a relative measurement of the $\bar{\nu}_{e}$ rate and spectrum between near and far ADs using 621 days of data\cite{man:dyb_621days}, which has 3.6 more times of the statistics than our previous published results\cite{man:dyb_217days}. The left plot of Figure \ref{fig:osc_rate} shows the ratio of the detected to expected $\bar{\nu}_{e}$ events that assuming no oscillation versus the effective baseline for each AD. The right plot, on top panel, of Figure \ref{fig:osc_rate} shows the comparision of the prompt spectra measured at far ADs with the expected spectrum at far ADs based on the measured spectrum at near ADs with and without oscillation interpretations. While the bottom panel plot shows the ratio of the measured prompt spectrum at far ADs to the weighted near ADs' spectrum without oscillation. From Figure \ref{fig:osc_rate} we can see that both the observed relative rate deficit and the relative spectrum distortion are highly consistent with the oscillation interpretation.


\begin{figure}[h]
\includegraphics[width=250pt]{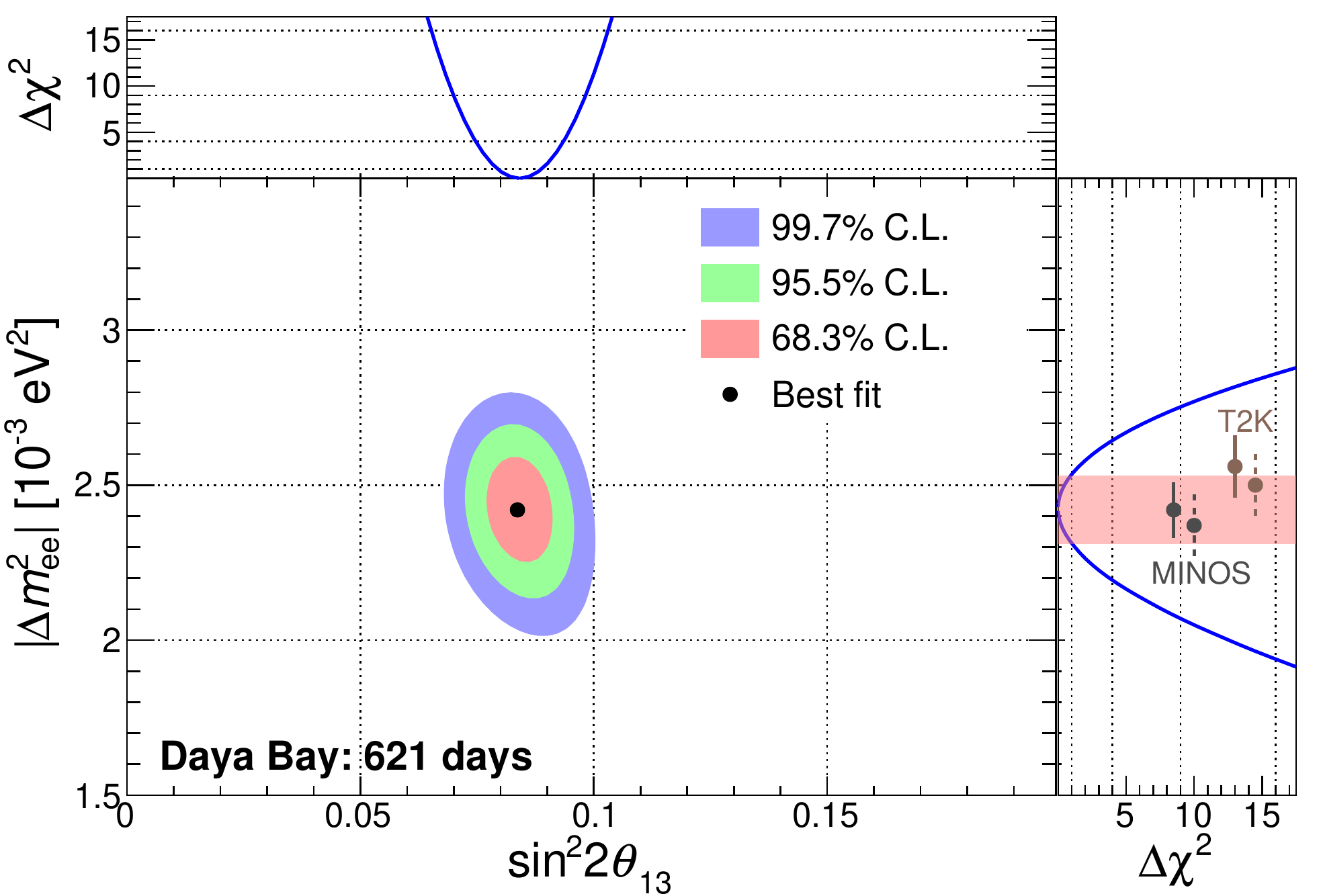}
\includegraphics[width=200pt]{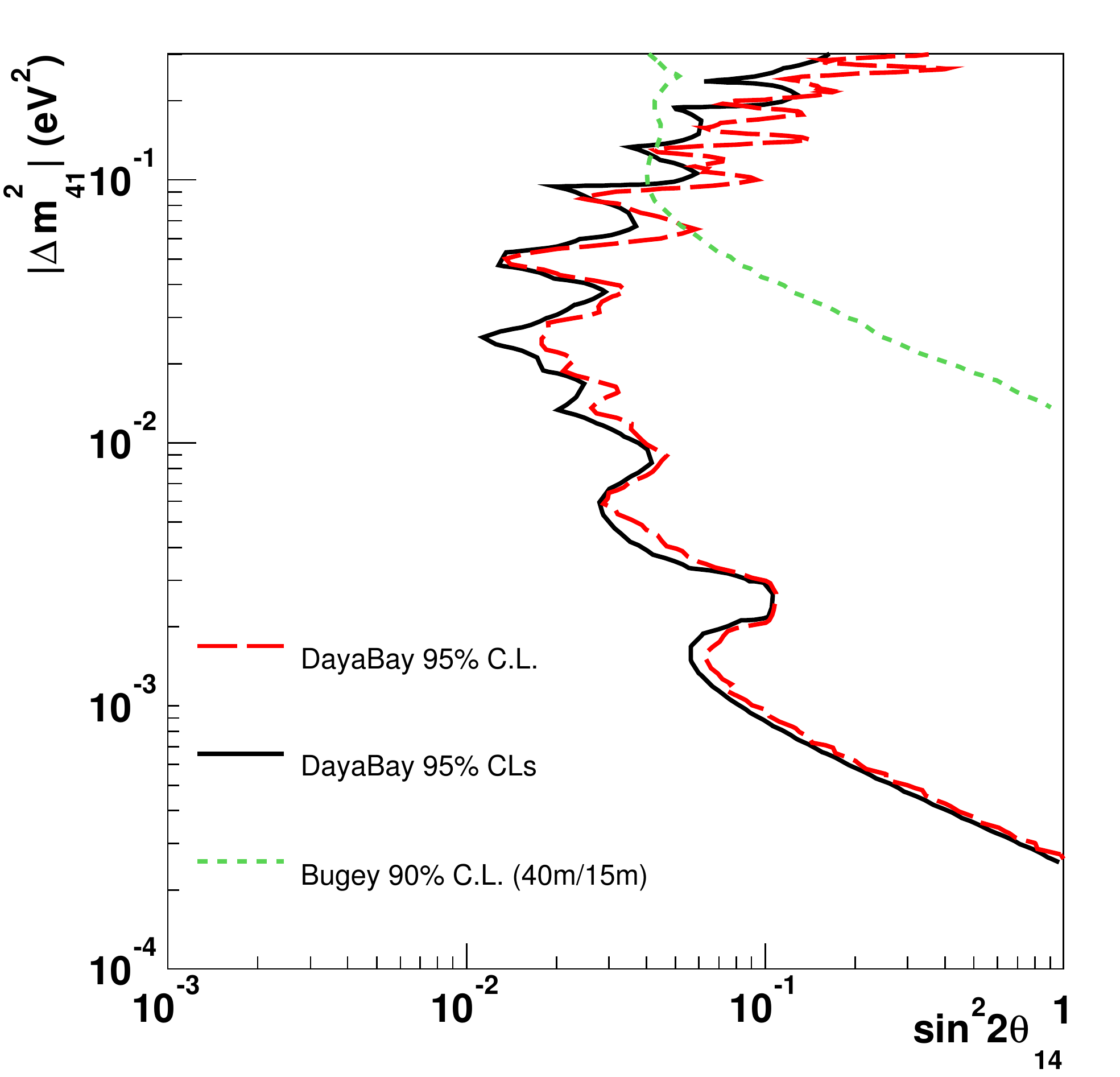}
\caption{(Color online) Left: The best fit value (black dot) of sin$^{2}2\theta_{13}$ and $|\Delta{}m_{ee}^{2}|$, which gives $\sin^{2}2\theta_{13}$ = 0.084 $\pm$ 0.005 and  $|\Delta{}m_{ee}^{2}|$ = 2.42 $\pm$ 0.11 $\times$ 10$^{-3}$ eV$^{2}$. The three bands surrounding the best fit value (from inner to outer) are corresponding to the allowed regions with 68.3\%, 95.5\% and 99.7\% confidence level (C.L.). The adjoint panels show the dependence of $\Delta{}\chi^{2}$ on sin$^{2}2\theta_{13}$ (top) and $|\Delta{}m_{ee}^{2}|$ (right). The band on the right adjoint panel shows the 1 $\sigma$ variation of $|\Delta{}m_{ee}^{2}|$. The solid and dashed dots are the converted $|\Delta{}m_{ee}^{2}|$ from $|\Delta{}m_{32}^{2}|$, from MINOS\cite{man:minos} and T2K\cite{man:t2k} experiments, assuming normal and inverted mass hierarchy, respectively. Right: The 90\% C.L. exclusion contours extracted from Daya Bay using the CLs method (solid) and the Feldman-Cousin method (long-dashed), the top short-dashed contour represents Bugey-3's 90\% C.L. exclusion contour using Raster scan method\cite{man:bugey}. }
\label{fig:osc_fit}
\end{figure}

The left plot of Figure \ref{fig:osc_fit} shows the best fit values of sin$^{2}2\theta_{13}$ and $|\Delta{}m_{ee}^{2}|$ from our oscillation analysis, which yields $\sin^{2}2\theta_{13}$ = 0.084 $\pm$ 0.005 and $|\Delta{}m_{ee}^{2}|$ = (2.42 $\pm$ 0.11) $\times$ 10$^{-3}$ eV$^{2}$. The precision of the sin$^{2}2\theta_{13}$ reaches 6\% and is currently the most precise measurement of this parameter in the world. The $|\Delta{}m_{ee}^{2}|$ measurement, with a hight precision of $\sim$ 5\%, is consistent with the muon neutrino disappearance measurement of $|\Delta{}m_{32}^{2}|$ by MINOS\cite{man:minos} and T2K\cite{man:t2k} experiments, converted to $|\Delta{}m_{ee}^{2}|$. More information of this Daya Bay oscillation analysis with 621 days of data can be found in Reference \cite{man:dyb_621days}. Daya Bay also performed an independent relative measurement of $\bar{\nu}_{e}$  rate  between near and far ADs of $\sin^{2}2\theta_{13}$ by detecting the IBD neutron captured on H (nH rate analysis). This analysis uses 217 days of data that with 6-AD configuration, and yields the $\sin^{2}2\theta_{13}$ value of 0.083 $\pm$ 0.018. The $\sin^{2}2\theta_{13}$ from nH rate analysis is also a very precise measurement of this parameter and it is consistent with the result from the relative rate and spectrum measurement of IBD neutron captured on Gd\cite{man:dyb_621days}. More information of the nH rate analysis with 217 days of data can be found in Reference \cite{man:dyb_nh}.


The oscillation analysis is performed in the 3 neutrino framework, what if there exists a fourth light sterile neutrino (with mass of m$_{4}$), will this affect our oscillation results? To answer these questions, Daya Bay also performed the light sterile neutrino search in the 3 (active) + 1 (sterile) neutrino framework using 217 days of data\cite{man:dyb_sterile}. No significant signal had been observed for the existence of a light sterile neutrino. However, the Daya Bay is able to set the most stringent limit on $\sin^{2}2\theta_{14}$ in the $|\Delta{}m_{41}^{2}|$ region from 10$^{-3}$ to 10$^{-1}$ eV$^{2}$. The right plot of Figure \ref{fig:osc_fit} shows the 90\% C.L. exclusion contours extracted from Daya Bay using the CLs method (solid) and the Feldman-Cousin method (long-dashed), while the top short-dashed contour represents Bugey-3's 90\% C.L. exclusion contour using Raster scan method\cite{man:bugey}.

\begin{figure}[h]
\includegraphics[width=250pt]{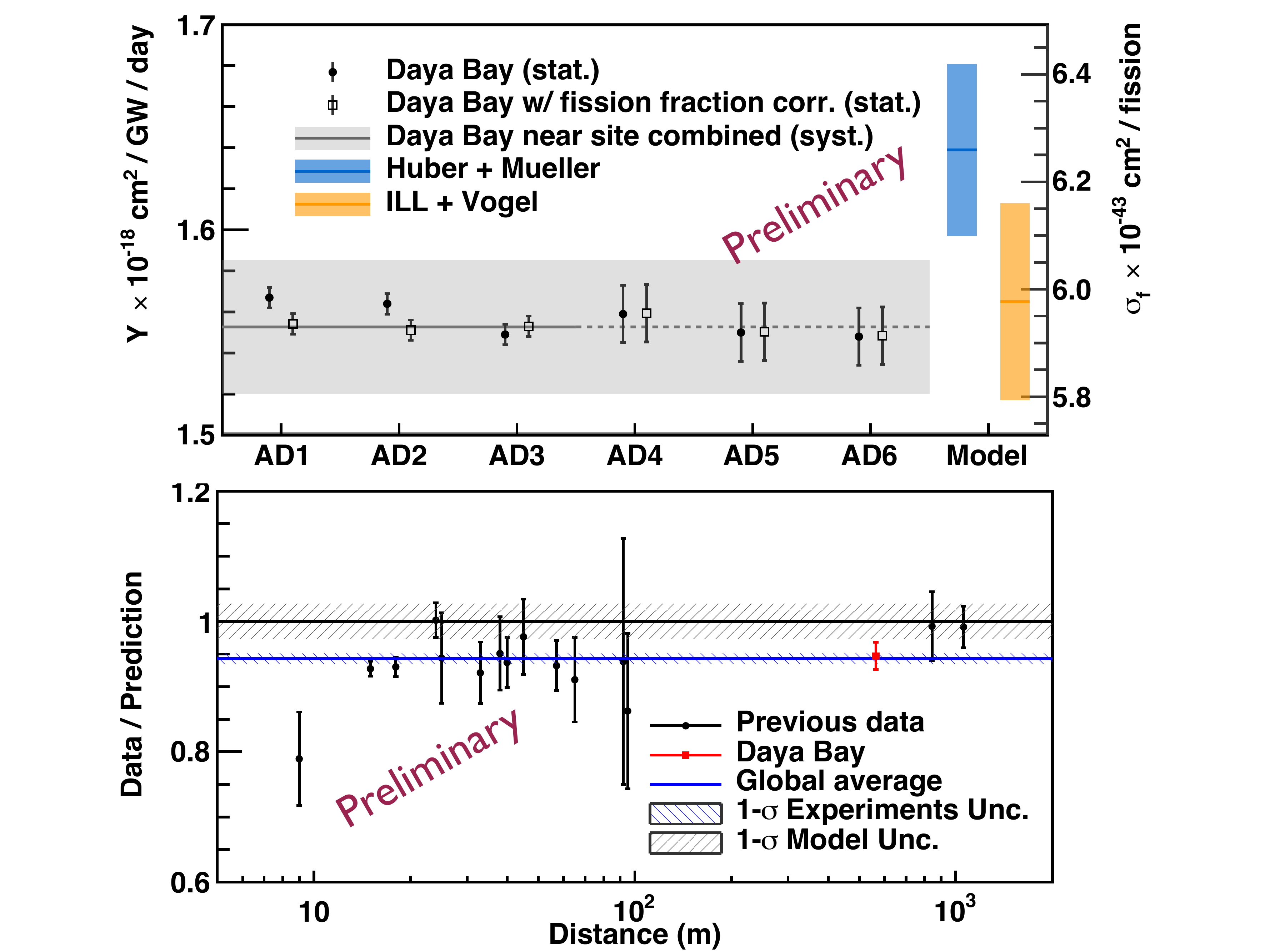}
\includegraphics[width=255pt]{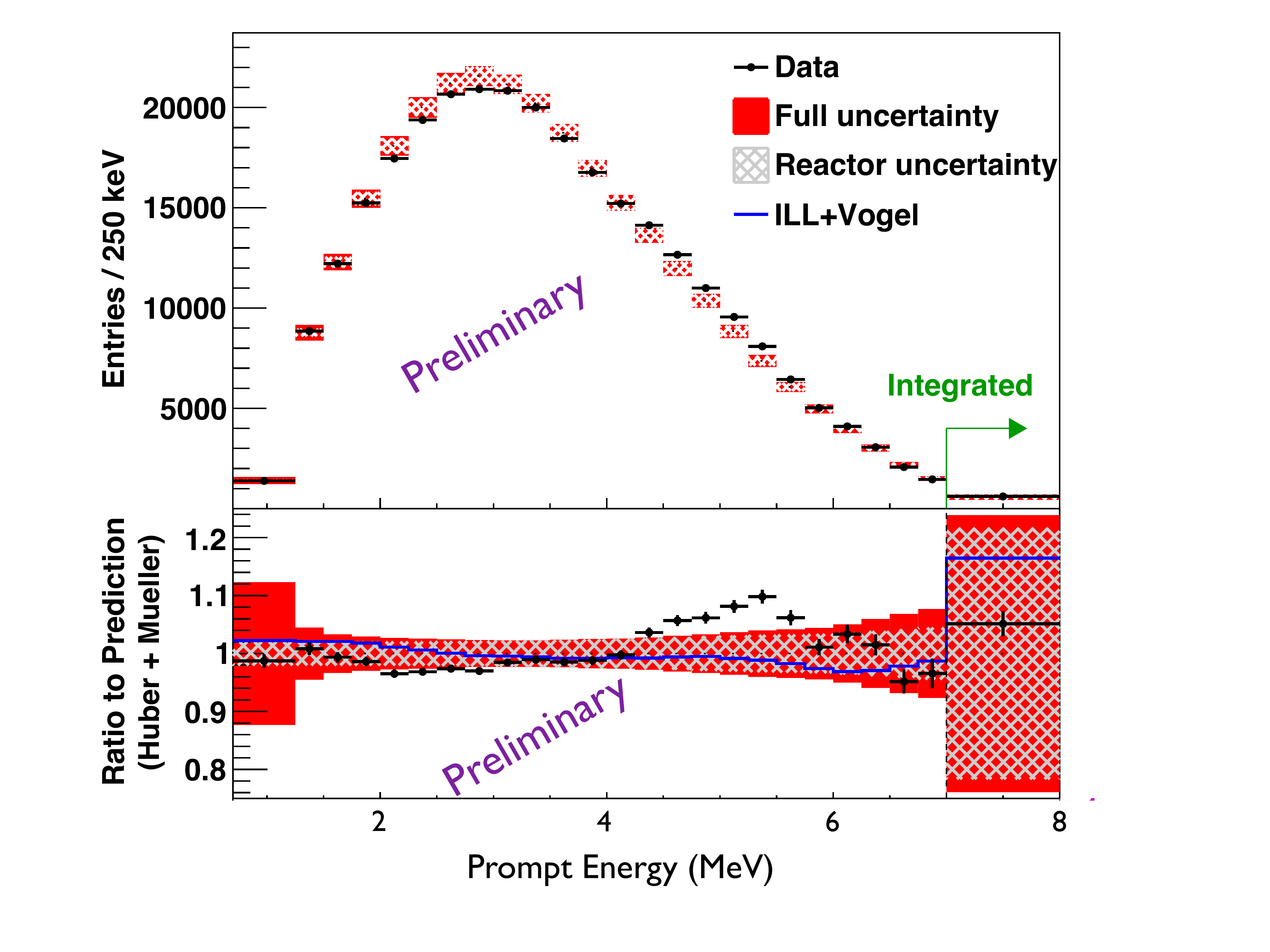}
\caption{(Color online) Left: (Top Panel) Rate of reactor $\bar{\nu_{e}}$ candidate events in the 6 ADs with corrections for 3-flavor oscillations (closed circles), and additionally for the variation of flux-weighted fission fractions at the different experimental halls (open squares). The average of the three near ADs is shown as the straight line (extended to the far ADs) with 1 systematic uncertainty band. The rate predicted with the Huber+Mueller\cite{man:huber, man:mueller} (ILL+Vogel\cite{man:ill, man:vogel}) models and its uncertainty are shown at the right. (Bottom Panel) The measured reactor $\bar{\nu}_{e}$ rate as a function of the distance from the reactor, normalized to the theoretical prediction with the Huber+Mueller model. The rate is corrected for 3-flavor neutrino oscillations at each baseline. The lower shaded region represents the global average and its 1 $\sigma$ uncertainty. The 2.7\% model uncertainty is shown as a band around unity. Measurements at the same baseline are combined for clarity. The Daya Bay measurement is shown at the flux-weighted baseline (573 m) of the two near halls. Right: Top panel: Predicted and measured prompt-energy spectra. The prediction is based on the Huber+Mueller model and normalized to the number of measured events. The highest energy bin contains all events above 7 MeV. The error bars on the data points represent the statistical uncertainty. Bottom panel: Ratio of the measured prompt-energy spectrum to the predicted spectrum (Huber+Mueller model). The curve shows the ratio of the prediction based on the ILL+Vogel model to that based on the Huber+Mueller model.}
\label{fig:absolute}
\end{figure}

The absolute reactor $\bar{\nu}_{e}$ flux and spectrum measurements were also performed at Daya Bay. With 217 days of data, A total of 300k and 40k IBD candidates are detected at the near and far ADs, respectively. The measured $\bar{\nu}_{e}$ flux is (1.55 $\pm$ 0.04) $\times$ 10$^{-18}$ cm$^{2}$/GW/day or (5.92 $\pm$ 0.14) $\times$ 10$^{-43}$ cm$^{2}$/fission (see the top panel plot on the left side of Figure \ref{fig:absolute}), and it is consistent with previous short baseline reactor $\bar{\nu}_{e}$ mesurements. The ratio of the measured $\bar{\nu}_{e}$ flux to the predcited flux from Huber+Mueller reactor $\bar{\nu}_{e}$ models\cite{man:huber, man:mueller} is 0.946 $\pm$ 0.022, which is consistent with the world average ratio of 0.942 $\pm$ 0.009 (see the bottom panel plot on the left side of Figure \ref{fig:absolute}).     

The plot at the top panel on the right side of Figure \ref{fig:absolute} shows the comparison of the absolute reactor spectrum measured to the prediction with Huber+Mueller reactor $\bar{\nu}_{e}$ model. For the whole energy range, there is a 2.6 $\sigma$ discrepancy between the measured and predicted spectra. An excess structure has been observed in the prompt energy range from 4 to 6 MeV, the discrepancy in this region yields a 4 $\sigma$ (see the plot at the bottom panel on the right side of Figure \ref{fig:absolute}). From the right plots of Figure \ref{fig:absolute}, we can see that the measured absolute reactor $\bar{\nu}_{e}$ spectrum is not consistent with the prediction of all reactor $\bar{\nu}_{e}$ flux models. However, this discrepancy has no impact on the sin$^{2}2\theta_{13}$ and $|\Delta{}m_{ee}^{2}|$ values, since they are extracted from the relative measurement of rate and spectrum between near and far ADs.

\section{SUMMARY}

The Daya Bay neutrino experiment has measured $\sin^{2}2\theta_{13}$ = 0.084 $\pm$ 0.005 and $|\Delta{}m^{2}_{ee}|$ = 2.42 $\pm$ 0.11 $\times$ 10$^{-3}$ eV$^{2}$ using 621 days of data. This measurement is currently the most precise measurement of $\sin^{2}2\theta_{13}$ in the world. The $|\Delta{}m^{2}_{ee}|$ measurement also has a comparable precision and is consistent with the muon disappearance experiment results from MINOS and T2K experiments. The light sterile neutrino searching analysis is able to set the most stringent limit on $\sin^{2}2\theta_{14}$ in the $|\Delta{}m^{2}_{14}|$ region from 10$^{3}$ to 10$^{-1}$ eV$^{2}$. The absolute reactor $\bar{\nu}_{e}$ flux measurement is consistent with previous short-baseline reactor neutrino experiments. While the absolute reactor $\bar{\nu}_{e}$ spectrum is not consistent with predictions from all models. Daya Bay will keep running until the end of 2017, by that time, both $\sin^{2}2\theta_{13}$ and $|\Delta{}m^{2}_{ee}|$ precisions are expected to reach 3\%.




\section{ACKNOWLEDGMENTS}
Daya Bay is supported in part by the Ministry of Science and Technology of China, the U.S. Department of Energy,
the Chinese Academy of Sciences, the National Natural Science Foundation of China, the Guangdong provincial gov-
ernment, the Shenzhen municipal government, the China General Nuclear Power Group, Key Laboratory of Particle
and Radiation Imaging (Tsinghua University), the Ministry of Education, Key Laboratory of Particle Physics and
Particle Irradiation (Shandong University), the Ministry of Education, Shanghai Laboratory for Particle Physics and
Cosmology, the Research Grants Council of the Hong Kong Special Administrative Region of China, the University
Development Fund of The University of Hong Kong, the MOE program for Research of Excellence at National Taiwan
University, National Chiao-Tung University, and NSC fund support from Taiwan, the U.S. National Science Founda-
tion, the Alfred P. Sloan Foundation, the Ministry of Education, Youth, and Sports of the Czech Republic, the Joint
Institute of Nuclear Research in Dubna, Russia, the CNFC-RFBR joint research program, the National Commission of
Scientific and Technological Research of Chile, and the Tsinghua University Initiative Scientific Research Program.
We acknowledge Yellow River Engineering Consulting Co., Ltd., and China Railway 15th Bureau Group Co., Ltd.,
for building the underground laboratory. We are grateful for the ongoing cooperation from the China General Nuclear
Power Group and China Light and Power Company.



\end{document}